\documentclass{vldb}
\usepackage{mathtools}
\usepackage{amsmath}
\usepackage{graphicx}
\usepackage{balance}  % for  \balance command ON LAST PAGE  (only there!)
\usepackage{caption}

\begin{document}

% ****************** TITLE ****************************************

\title{Join Query Optimization with Deep Reinforcement Learning Algorithms}

% possible, but not really needed or used for PVLDB:
%\subtitle{[Extended Abstract]
%\titlenote{A full version of this paper is available as\textit{Author's Guide to Preparing ACM SIG Proceedings Using \LaTeX$2_\epsilon$\ and BibTeX} at \texttt{www.acm.org/eaddress.htm}}}

% ****************** AUTHORS **************************************


\numberofauthors{2} %  in this sample file, there are a *total*

\author{
% 1st. author
\alignauthor
Jonas Heitz\\
       \affaddr{Zurich University of Applied Sciences}\\
       \affaddr{Switzerland}
       %\affaddr{Obere Kirchgasse 2}\\
       %\affaddr{Winterthur, Switzerland}
        \email{J.Heitz@bluewin.ch}
% 2nd. author
\alignauthor
Kurt Stockinger\\
       \affaddr{Zurich University of Applied Sciences}\\
       \affaddr{Switzerland}
       %\affaddr{Obere Kirchgasse 2}\\
       %\affaddr{Winterthur, Switzerland}
       \email{Kurt.Stockinger@zhaw.ch}
%
% 3rd. author
%\alignauthor Lars Th{\Large{\sf{\o}}}rv{$\ddot{\mbox{a}}$}ld\titlenote{This author %is the
%one who did all the really hard work.}\\
%       \affaddr{The Th{\large{\sf{\o}}}rv{$\ddot{\mbox{a}}$}ld Group}\\
%       \affaddr{1 Th{\large{\sf{\o}}}rv{$\ddot{\mbox{a}}$}ld Circle}\\
%       \affaddr{Hekla, Iceland}\\
%       \email{larst@affiliation.org}
%\and  % use '\and' if you need 'another row' of author names
% 4th. author
%\alignauthor Lawrence P. Leipuner\\
%       \affaddr{Brookhaven Laboratories}\\
%       \affaddr{Brookhaven National Lab}\\
%       \affaddr{P.O. Box 5000}\\
%       \email{lleipuner@researchlabs.org}
%% 5th. author
%\alignauthor Sean Fogarty\\
%       \affaddr{NASA Ames Research Center}\\
%%       \affaddr{California 94035}\\
%       \email{fogartys@amesres.org}
%% 6th. author
%\alignauthor Charles Palmer\\
%       \affaddr{Palmer Research Laboratories}\\
%       \affaddr{8600 Datapoint Drive}\\
%       \affaddr{San Antonio, Texas 78229}\\
%       \email{cpalmer@prl.com}
}

% There's nothing stopping you putting the seventh, eighth, etc.
% author on the opening page (as the 'third row') but we ask,
% for aesthetic reasons that you place these 'additional authors'
% in the \additional authors block, viz.
%\additionalauthors{Additional authors: John Smith (The Th{\o}rv\"{a}ld Group, %{\texttt{jsmith@affiliation.org}}), Julius P.~Kumquat
%(The \raggedright{Kumquat} Consortium, {\small \texttt{jpkumquat@consortium.net}}), %and Ahmet Sacan (Drexel University, {\small \texttt{ahmetdevel@gmail.com}})}
%\date{30 June 2019}
% Just remember to make sure that the TOTAL number of authors
% is the number that will appear on the first page PLUS the
% number that will appear in the \additionalauthors section.

\maketitle

\begin{abstract}
Join query optimization is a complex task and is central to the performance of query processing. In fact it belongs to the class of NP-hard problems. Traditional query optimizers use dynamic programming (DP) methods combined with a set of rules and restrictions to avoid exhaustive enumeration of all possible join orders. However, DP methods are very resource intensive. Moreover, given simplifying assumptions of attribute independence, traditional query optimizers rely on erroneous cost estimations, which can lead to suboptimal query plans.  

Recent success of deep reinforcement learning (DRL) creates new opportunities for the field of query optimization to tackle the above-mentioned problems. In this paper, we present our DRL-based \underline{F}ully \underline{O}bserved \underline{Op}timizer (FOOP) which is a generic query optimization framework that enables plugging in different machine learning algorithms. The main idea of FOOP is to use a data-adaptive learning query optimizer that avoids exhaustive enumerations of join orders and is thus significantly faster than traditional approaches based on dynamic programming. In particular, we evaluate various DRL-algorithms and show that Proximal Policy Optimization significantly outperforms Q-learning based algorithms. Finally we demonstrate how ensemble learning techniques combined with DRL can further improve the query optimizer.
\end{abstract}

%% Your real content!
\section{Introduction}
\label{sec:intro}
Query optimization has been studied over decades and is one of the key aspects in database management systems (DBMS). However, one of the biggest challenges in query optimization is {\it join order optimization}, i.e. to find the optimal order of executing joins such that the query costs and hence the query execution time is minimized \cite{marcus2018deep}. Since join order optimization belongs to the class of NP-hard problems \cite{leis2015good}, exhaustive query plan enumeration is a prohibitive task for large databases with multi-way join queries. Hence, query optimization is often considered as a trade-off between the quality of a query plan and the time spent optimizing it.

Query optimization strategies of many commercial DBMSes are based on ideas introduced in System R \cite{astrahan1976system} or in the Volcano Optimizer Generator \cite{graefe1991volcano}. These systems use dynamic programming (DP) combined with a set of rules to find good query plans. These rules prune the search space of potential query plans, which reduces the time spent optimizing the query but also lowers the chance that the optimal query plan is found in the large search space. Traditional query optimizers suffer from a second issue besides the limitation of the search strategies. They rely on cost models to estimate the cost of executing a query. These cost models are built on cardinality estimations which are based on quantile statistics, frequency analysis or even non-theoretically grounded methods \cite{leis2015good}. Errors in the cardinality estimation often lead to suboptimal query plans. Moreover, traditional query optimizers do not learn from prior executed queries. Even though concepts of learning optimizers are around since LEO \cite{stillger2001leo}, these approaches have not been widely adopted. As the work of Leis et al. \cite{leis2015good} shows, there is a need for data-adaptive learning query optimizers for large analytical databases.

Recent success in deep reinforcement learning (DRL) has brought new opportunities to the field of query optimization. For example, ReJoin \cite{marcus2018deep} and DQ \cite{krishnan2018learning} propose their approaches to use DRL to optimize join queries. Both papers apply different DRL algorithms in their query optimizers. However, there is no generic query optimization framework which allows studying different machine learning algorithms and hence enables a direct comparison between different methods.

In this paper we introduce a DRL-based \underline{F}ully \underline{O}bserved \underline{Op}timizer (FOOP) for databases. FOOP allows  reinforcement learning (RL) algorithms to track all relations and intermediate results during the query optimization process, similar to the observations during a game of Go \cite{silver2016mastering}. FOOP is inspired by ReJoin \cite{marcus2018deep} and DQ \cite{krishnan2018learning} and enhances them with the most recent modules introduced in DLR research like Double Deep Q-Networks and Priority Replay. We show that FOOP produces query plans with similar quality to traditional query optimizers using significantly less time  for optimization.\

\pagebreak

The paper makes the following contributions:

\begin{itemize}
\item  With FOOP we introduce query optimization as a fully observable RL problem, which allows RL algorithms to track all relations and intermediate results during the optimization process. We place FOOP in the application layer, which makes the optimizer DBMS independent.
\item To the best of our knowledge, this paper presents the first face-to-face comparison of DRL algorithms vanilla Deep Q-Network, Double Deep Q-Network with Priority Replay and Proximal Policy Optimization in query optimization.
%\item The basic requirements for RL algorithms to learn database structures are evaluated.
\item FOOP produces query plans with cost estimations in a similar range to traditional query optimizers by using optimization algorithms with significantly lower runtime complexity.
\end{itemize}

The paper is organized as follows. Section \ref{sec:relwork} reviews the literature in the areas of query optimization. Section \ref{sec:bg} provides the background knowledge about query optimization and the basics of reinforcement learning. Section \ref{sec:arch} introduces the architecture and the featurization of our learning optimizer FOOP. Section \ref{sec:eval} provides a detailed evaluation of the experiments with FOOP. Finally, we present our conclusions and future work in Section \ref{sec:conc}.
\section{Related Work}
\label{sec:relwork}
Query optimization is a well-studied problem in the data- base community with many different solutions proposed over the last decades. Pioneering work dates back to static query optimization of System R \cite{astrahan1976system} and the Vulcano Optimizer Generator \cite{graefe1991volcano}, which has been widely used in commercial systems \cite{hellerstein2005readings}. Later, researchers introduced new architectures for query optimization, where  queries are continuously optimized and validated during query processing \cite{avnur2000eddies} \cite{markl2004robust}. In 2001 IBM introduced the learning optimizer LEO for DB2, which is based on the architecture of static query optimization and is able to learn from its mistakes \cite{stillger2001leo}.

Lohman states in his blog post \cite{lohman2014query} that query optimization is still a unsolved problem and pointed out that most query benchmarks used in research do not reflect databases in the real world. Leis et al. picked up on that thought and created a new query benchmark to demonstrate the issues of query optimizers in commercial DBMSes \cite{leis2015good}. 

Recent progress in machine learning established new ways to tackle those open problems. For instance, some approaches try to mitigate the issue of cardinality estimation errors by introducing machine learning models to predict the costs or the execution time of queries \cite{yusufoglu2014neural} \cite{wu2013towards} \cite{akdere2012learning} \cite{ganapathi2009predicting} \cite{li2012robust} \cite{marcus2016wisedb}. 

Others apply more modern machine learning approaches leveraging recent advanced in reinforcement learning \cite{wang2016database} \cite{marcus2018towards}. The main idea of \cite{marcus2018towards} is to automatically tune a particular database and to improve the performance of query execution by using the feedback from past query executions. Oritz et al. studied how state representation affects query optimization when using reinforcement learning  \cite{ortiz2018learning} for static query optimization. The Skinner DB system uses reinforcement learning for continuous query optimization \cite{trummer2018skinnerdb}. In that approach, queries are dynamically improved on the base of a regret bounded quality measure. Most recently Marcus et al. introduced the end-to-end learning optimizer Neo \cite{marcus2019neo} which is inspired by AlphaGo \cite{silver2016mastering}. Neo uses a learned cost model based on neuoral networks (NN) to guide a search algorithm through the large search space of all possible query plans. The approach was motivated by the NN-guided Monte Carlo Tree Search which got famous through AlphaGoZero. 

For our paper, we focus on Q-learning and policy gradient based reinforcement learning, which got popular after the publication of the Atari paper by DeepMind \cite{mnih2013playing}. The closest works to ours are ReJoin \cite{marcus2018deep} and DQ \cite{krishnan2018learning}. ReJoin uses policy gradient methods to find a good policy for join order enumeration, whereas DQ focus on a Deep Q-Network architecture to optimize their query plans. Both papers rely on traditional cost models to  find optimal query plans -- which is suboptimal in terms of erroneous cardinality estimations. However, DQ introduces a mechanism to fine-tune their Deep Q-Network on the query execution times to mitigate this issue. The advantage of Deep-Q-Networks and policy gradient methods over the NN-guided plan search of Neo is the shorter training time. 

The key advantage of our approach over all other approaches is that we provide a generalized architecture, which allows to use different front line RL algorithms with minimal effort. FOOP places the query optimizer in the application layer which makes the approach DBMS independent. Further, we introduce query optimization as a fully observed RL problem, which allows the RL algorithms to learn new aspects about the query execution performance which might be difficult to achieve with traditional cost-based optimizers.

\section{Background}
\label{sec:bg}
In this section, we provide background knowledge about query optimization and reinforcement learning (RL). First, we elaborate why query optimization is still an unsolved problem and its main issues. Later we give a brief introduction into RL and describe the methods we use for our learning query optimizer.

\subsection{Query Optimization}
\label{sec:bg_qo}
The goal of a query optimizer in a database system is to translate queries, which are written in a declarative language (e.g. SQL), into an efficient query execution plan. %First, we discuss the complexity of the query optimization problem. Then we introduce the standard components of a query optimizer, to show how traditional query optimizers work. Lastly, we discuss the issues, which evolve from traditional query optimizers.
\subsubsection{Complexity}
\label{sec:bg_qo_comp}
To understand the complexity of query optimization, we look at the join optimization problem, a sub problem of query optimization in database systems. Join order optimization is the task of finding the best arrangement of relations (tables) in a multi-way join such that the estimated execution cost is minimized. In other words, we try to find best query execution plan for a given query. 

One of the simplest solutions to solve this optimization problem is to only consider all possible permutations of those relations that are involved in the join. In this case, every (join) query with $n$ relations can be expressed with $n!$ different join paths. For instance, the query '$A \bowtie B \bowtie C$' could be executed either as '$A \bowtie B \bowtie C$', or '$C \bowtie B \bowtie A$', or '$A \bowtie C \bowtie B$' etc. This problem has the complexity $O(n!)$.

%In  Figure \ref{fig:b_qo_permuatuions} you can see all permutations of our sample query introduced in Appendix \ref{ap:db}.   

%\clearpage
%\begin{figure}[ht]
%\includegraphics[width=8.5cm]{imgs/bg_permutation.png}
%\caption{All permutations for a three-way-join-query (P refers to product, %OI to orderItem, O to order and C to customer)}
%\label{fig:b_qo_permuatuions}
%\end{figure}

The simplifying assumption for this complexity estimation is to not distinguish between the left-side and the right-side of the join operators. However, a query optimizer also needs to take into account aspects of physical query optimization. For instance, a join could be executed as a nested loop join, a hash join, a sort-merge join, etc. As a consequence, the complexity of query optimization grows rapidly. According to the principal of Catalan Numbers \cite{crepinvsek2009efficient}, the number of all possible binary (query) tree shapes for a j-way-join-query is $(2j)! / (j! * (j+1)!)$ \cite{crepinvsek2009efficient}, where a j-way-join has $n$ relations ($j=n-1$). 

As a running example for this paper, assume a small database with the following schema depicted in Figure \ref{fig:sample_db_er}:

\begin{figure}[ht]
\includegraphics[width=8.5cm]{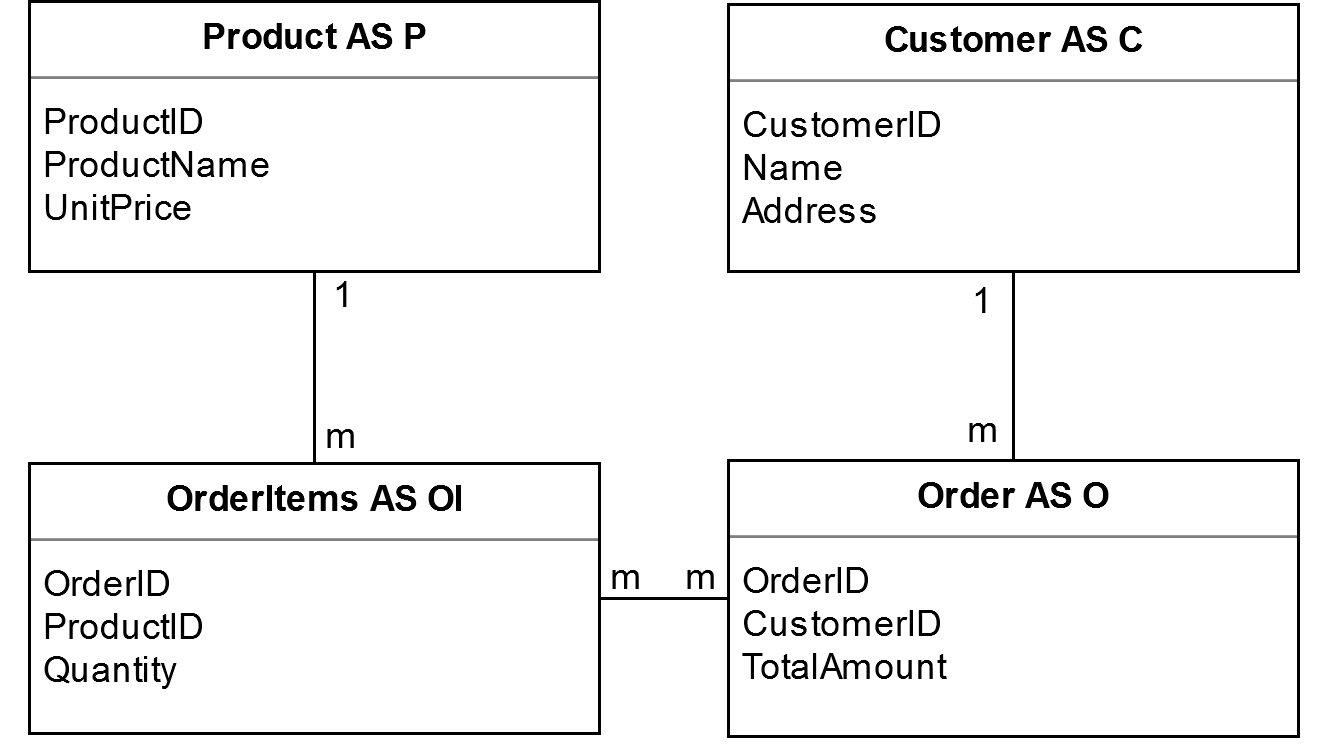}
\caption{Sample database}
\label{fig:sample_db_er}
\end{figure}

\begin{figure}[ht]
\includegraphics[width=8.5cm]{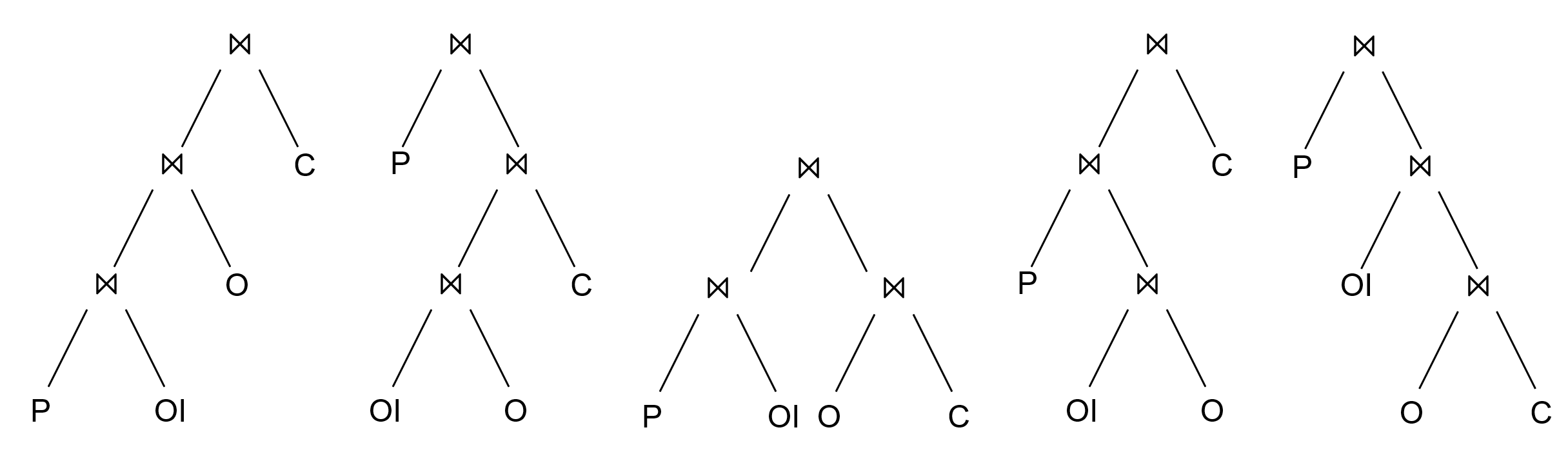}
\caption{Some possible binary-tree shapes for a three-way-join-query (P refers to product, OI to orderItem, O to order and C to customer).}
\label{fig:b_qo_trees}
\end{figure}

Further assume that we want to optimize the following query: '$P \bowtie OI \bowtie O \bowtie C$'. Let us now analyze all possible query plans for the above-mentioned query. Figure \ref{fig:b_qo_trees} shows some binary tree shapes (query execution plans) of our three-way-join-query.  

%Multiplied with all permutations we get the complexity class of \begin{equation}
%O(n!* (2(n-1))! / ((n-1)! * n!)) 
%\end{equation}
%respectively 
%\begin{equation}
%O( (2n)!/(n-1)!).
%\end{equation}  

The complexity of the problem even rises, if we consider multiple different join algorithms or additional methods like selections, projections, etc. The query optimization literature often uses just the lower bound complexity of $\Omega(n!)$. However, finding an optimal query plan in a large search space of $\Omega(n!)$ is daunting task. Moreover, since join operations of relational database systems are commutative and associative, it is not surprising that query optimization belongs to the class of NP-hard problems \cite{ioannidis1996query}.

\subsubsection{Components}
\label{sec:bg_qo_compo}
In this subsection we look into the key components of a query optimizer. Many DBMSes follow the traditional textbook architecture of System R \cite{astrahan1976system} and the Volcano Optimizer Generator \cite{graefe1991volcano}. The architecture of these query optimizers is modular and consists of at least two components that we will analyze below. Moreover, we use PostgreSQL \cite{postgresql2019postgresql}, as a representative of widely used open source DBMS, to show how those modules are implemented. 

\begin{itemize}
\item \textbf{Query planner}: The planner (also called plan enumerator) is the central module of the optimizer. It examines various possible query plans and chooses the optimal one based on a certain cost model. However, since exhaustive enumeration of all possible query plans is often computationally too expensive \cite{vt22019vt2}, query planers typically use heuristics to reduce the search complexity. For instance, System R uses dynamic programming combined with a set of rules, to reduce the search space \cite{ioannidis1996query}. Also PostgreSQL uses a dynamic programming algorithm but switches to a greedy or genetic approach if a query exceeds a certain size \cite{leis2015good}. To reduce the complexity of the problem, traditional query optimizers typically only consider {\it left-deep query trees} and prohibit Cartesian products \cite{ioannidis1996query}.

\item \textbf{Cost model}: The cost model delivers the cost estimations for a query plan according to different assumptions and statistics about the database \cite{vt22019vt2}. In the case of PostgreSQL, the cost model uses a complex combination of estimated CPU-, I/O-costs and cardinality estimations \cite{leis2015good}. Further more, the cardinality is predicted in simple cases (e.g. for the base table) with quantile statistics, frequency analysis or distinct counts. For complex predicates, PostgreSQL uses non-theoretically grounded constants \cite{leis2015good}.
\end{itemize}

\subsubsection{Open Issues}
\label{sec:bg_qo_oi}
There are multiple unsolved issues in query optimization. In the following we present three of them:

\begin{itemize}
\item \textbf{Query planner limitations}: Many query optimizers only use left-deep query trees, since exhaustive enumeration would be too time consuming. The target of this constraint is to lower the complexity of the problem to $O(n!)$ \cite{ioannidis1996query}. However, this approach reduces the chance that the optimizer finds the optimal query plan. Usually the best left-deep query tree is close to the optimal query plan. Unfortunately, that is not true for large analytical databases in combination with complex queries, where this search space reduction can have time consuming consequences during query execution \cite{leis2015good}. 
\item \textbf{Query optimization benchmarks}: Traditional query optimizers do not work well with databases where the values between different columns are correlated or tables contain data following non-uniform distributions. Real world data sets are usually non-uniformly distributed and are highly correlated, as Leis et al. \cite{leis2015good} pointed out. Further, they showed that the benchmarks TPC-H, TPC-DS and the Star Schema Benchmark (SSB), which are mainly used to compare and evaluate query optimizers, are based on data sets with uniform distribution and independent columns \cite{leis2015good}. This means that traditional query optimizers, that are tuned based on these benchmarks, are optimized on a corner case, which does not represent the majority of real world databases \cite{leis2015good}.
\item \textbf{Estimation errors}: Further quantile statistics can be misleading on data sets with a non-uniform distribution and the correlated columns cannot get detected since the various columns do not get compared in traditional query optimizers. These issues result in estimation errors from the cost model. Faulty cost estimations lead the planner on a wrong track to suboptimal query plans \cite{vt22019vt2}.
\end{itemize}

\subsection{Reinforcement Learning (RL)}
\label{sec:bg_rl}
Reinforcement learning can be considered as  "learning what to do" with the goal to map situations to actions \cite{sutton2018reinforcement} -- which gets close to our understanding of the nature of learning. Consider, for instance, an infant waving its arms and looking around trying to interact with the environment. At this stage the infant does not have a teacher, it just produces information about cause and effect, which action has which consequence \cite{sutton2018reinforcement}. 

When we look at this paradigm from a computer science perspective, we can interpret it as a function. The function tries to maximize a reward signal, without being told which action it has to take \cite{sutton2018reinforcement}. Bellmann's 'Principle of Optimality' and the concept of dynamic programming laid the foundation for RL to be applied in computer science \cite{krishnan2018learning}. Dynamic programming has is already been used in query optimization for decades. Moreover, recent improvements in deep reinforcement learning seem to be a promising solution for query optimization. Hence, we will now revise the main concepts of reinforcement learning that we use for our query optimizer.

First we introduce the Markov Decision Process \cite{krishnan2018learning} as the foundation of RL. Afterwards  we discuss Q-learning and Policy Gradient Methods \cite{sutton2018reinforcement}.

\subsubsection{Markov Decision Process (MDP)}
\label{sec:bg_rl_mdp}

From a technical perspective, RL is a class of stochastic optimization methods that can be formulated as a Markov Decision Process (MDP) \cite{krishnan2018learning}. The basic idea is that an agent takes a sequence of actions with the goal to optimize a given objective in an MDP model. %However, MDPs are usually very difficult to solve exactly \cite{sutton2018reinforcement}. 

MDPs are an extension of Markov Chains and have to satisfy the Markov Property, i.e. future state progressions only depend on the current state and not on states and actions taken in the past. An MDP is therefore formalized as the following five-tuple \cite{krishnan2018learning}:

\begin{equation} 
\langle S,A,P(s,a),R(s,a),S_0 \rangle
\label{eq:mdp}
\end{equation}

\begin{itemize}
\item $S$ stands for {\it states} in which the agent can be.
\item $A$ stands for {\it actions} the agent can take to get to a new state.
\item $S_{t+1} \sim P(s,a)$ is the {\it new state} you reach through taking action \textit{a} in state \textit{s}. 
\item $R(s,a)$ stands for the {\it reward} the agent receives for being in state \textit{s} and taking action \textit{a}.
\item $S_0$ is the {\it initial state} the in which agent starts.

\end{itemize}

\begin{figure}[ht]
\includegraphics[width=8.5cm]{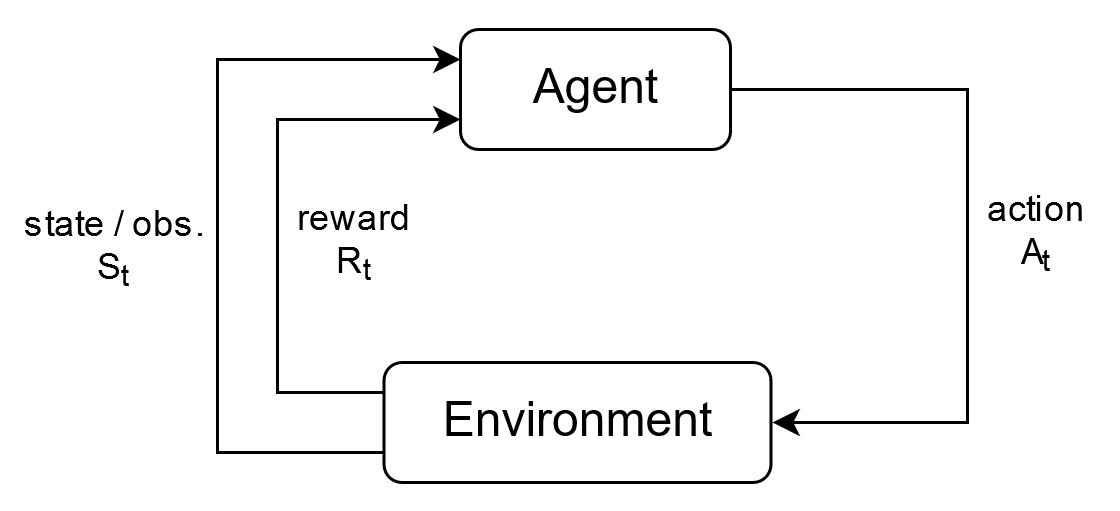}
\caption{Agent-environment interface for MDP interaction }
\label{fig:b_rl_arch}
\end{figure}

The overall goal is to find a decision policy $\pi: S \mapsto A$ (a function that maps states to actions), such that the expected reward is maximized \cite{krishnan2018learning}. Therefore, MDPs can be considered as a "learning from interactions". Figure \ref{fig:b_rl_arch} illustrates how these interactions work: The \textit{agent} is the learner, takes actions and interacts with the \textit{environment}. Moreover, the agent receives \textit{state} information and \textit{rewards} form the environment \cite{sutton2018reinforcement}. The state information is encoded and typically called \textit{observations}.

\subsubsection{Q-learning}
\label{sec:bg_rl_ql}

\textit{Temporal-Difference (TD)} learning is a central concept of RL. It combines the ideas of \textit{Monte Carlo (MC)} and \textit{Dynamic Programming (DP)}. TD learns directly from raw experience like MC and updates estimates without waiting for the final outcome like DP \cite{sutton2018reinforcement}. One of these TD control algorithms is called \textit{Q-learning}, that learns the {\it action-value function} $Q(s,a)$. This function tells the agent how good it is to take action \textit{a} while being in the state \textit{s}. 

Q-learning is an off-policy function, which means it is independent of the policy being followed. This simplifies the analysis of the algorithm drastically and allows fast convergence. Nevertheless, the current policy has an influence, since it determines which state-action pairs are visited and updated \cite{sutton2018reinforcement}. 

In traditional Q-learning there is a memory table which stores all Q-values \cite{sutton2018reinforcement}. However, typically there are too many combinations of states and actions to store and to compute in query optimization and other RL problems. To overcome that, we need to {\it approximate} the action-value function $Q(s,a)$. Any function approximator can be used, e.g. linear functions, non-linear functions, etc.

Neural networks (NN) have proven to be a good function approximator for non-linear functions \cite{goodfellow2016deep}. So it is not surprising that the combination of NNs and RL are the key idea behind the recent success of autonomously playing Atari \cite{mnih2015human} and Go \cite{silver2016mastering}. Relating to the Q-function approximation, Mnih et al. \cite{mnih2015human} introduced the \textit{Deep Q-Network} (DQN). The major challenges for using NNs in RL is that the inputs for the NN have to be independent \cite{goodfellow2016deep}. However, the input data is correlated by definition, since we collect experienced observations in RL during execution. To overcome this problem, Mnih et al. \cite{mnih2015human} proposed \textit{experience replay} and \textit{target network}:

\begin{itemize}
\item \textbf{Experience replay}: Experience replay is a buffer of historically gathered observation data from which the NN can randomly sample. This makes the training process more independent \cite{mnih2015human}.
\item \textbf{Target network}: There are two identical NNs created ($\theta^-,\theta$). The first NN is trained on all state-value updates. At certain intervals, the first NN sends the trained weights to the second NN, which does not require additional training. The second network can thus be used to fetch the Q-values.  This method helps to stabilize the training process \cite{mnih2015human}.
\end{itemize}

The problem can now be expressed with a maximization function shown in Equation \ref{eq:dqn}, which is optimized using stochastic gradient descent.

\begin{equation}
L =  R_{t+1} + \gamma_{t+1} max Q(S_{t+1},A_{t+1};\theta^-_t)-Q(S_t,A_t;\theta_t)^2
\label{eq:dqn}
\end{equation}

\begin{itemize}
\item $S_t, A_t, R_t$ are the state, action and reward at time $t$.
\item $\gamma$ is the discount factor for the long term reward.
\item $Q()$ is the Q-function and $\theta, \theta^-$ are the weights of the NN which approximates the Q-function.
\end{itemize}

The introduction of DQNs has been a major milestone in DRL. However, there are still several limitations. Hence, Hessel et al. \cite{hessel2018rainbow} summarized and combined some methods to improve the "vanilla" DQN. Some of their improvements are as follows:
\begin{itemize}
\item \textbf{Double DQN (DDQN)}: In standard DQNs, the learning process is affected by an overestimation bias, resulting from the maximization step in Equation \ref{eq:dqn}. DDQNs reduce the overestimation by using decoupling techniques, as shown in Equation \ref{eq:ddqn} \cite{hessel2018rainbow}.

\begin{equation}
\begin{aligned}
L = & R_{t+1} + \gamma Q(S_{t+1},max Q(S_{t+1},A_{t+1};\theta^-_t);\theta_{t}) - \\
    & Q(S_t,A_t;\theta_t)^2
\end{aligned}
\label{eq:ddqn}
\end{equation}

\item \textbf{Prioritized replay}: DQNs with experience replay sample uniformly from the replay buffer. 'Ideally, we want to sample those transitions, from which there is much to learn, more frequently' \cite{hessel2018rainbow}. Schaul et al. \cite{schaul2015prioritized} propose to sample transitions with the probability $p_t$, which is relative to the last encountered absolute \textit{TD error}, as a measurement for the learning potential:

\begin{equation}
p_t  \propto | r + \gamma max Q(S_{t+1},A_{t+1}’;\theta^-)- Q(S_t,A_t;\theta)|
\end{equation}

To bias the model towards recent transitions, the new inserted transitions receive maximum priority \cite{hessel2018rainbow}. 
\end{itemize}

\subsubsection{Policy Gradient Methods}
\label{sec:bg_rl_ppo}
So far we talked about \textit{action-value methods}, which learn values of actions and then select actions based on their estimated action value. \textit{Policy Gradient Methods} work differently. They learn a \textit{parameterized policy $\pi_\theta$} that selects actions without consulting a value function. 

Nevertheless, the value function is still used to learn the policy parameter \cite{sutton2018reinforcement}. $\theta$ represents the vector of the policy parameters. RL aims to identify the policy parameter $\theta$, which optimizes the expected reward $R_\pi(\theta)$. Unfortunately, it is not possible to compute $R_\pi(\theta)$ precisely, because that would mean to exhaustively compute every possible path. This is why \textit{gradient methods} search for the optimal policy parameters $\theta$ with an estimator $E$, i.e. the gradient of the reward ($E(\theta) \approx  \nabla_\theta R_\pi(\theta)$) \cite{marcus2018deep}. The policy $\pi_\theta$ can be represented as a NN, where the vector $\theta$ is used as the weights of the NN. This method is called {\it policy gradient deep learning} \cite{marcus2018deep}.

Many policy gradient methods are computationally too complex for real world tasks. The issue is due to the natural policy gradient, which involves a second-order derivative matrix. \textit{Proximal Policy Optimization (PPO)} introduced by Schulman et al. \cite{schulman2017proximal} is one of the policy gradient deep learning methods which uses a different approach. It realizes fast convergence and reliable performance from \textit{Trust Region Policy Optimization (TRPO)}, while using a first order optimizer like gradient descent. Similar to TRPO, PPO initializes a trust region in which the algorithm is allowed to look for a better policy \cite{schulman2017proximal}. In TRPO a surrogate objective function is defined, which is maximized to a constraint based on the size of the policy update. This problem can be approximately solved using the conjugate gradient algorithm after a linear approximation to the objective function and a quadratic approximation to the constraint \cite{schulman2017proximal}. In PPO, the constraint is formulized as a penalty in the objective function to avoid the quadratic approximation. 

A step further, PPO can also be implemented with the \textit{Clipped Surrogate Objective} function \cite{schulman2017proximal}.
\begin{equation}
r_t(\theta) = \frac{\pi_\theta(A_t|S_t)}{\pi_{\theta_{old}}(A_t|S_t)}\\
\end{equation}

In the equation above, $\pi_\theta$ is the new policy and $\pi_{\theta_{old}}$ is the old policy. The advantage of this function is that instead of penalising large policy changes, these changes are discouraged if they are "outside of the comfort zone" as shown in Equation \ref{eq:ppo_clip}. 

\begin{equation}
L^{CLIP} = E_t[min(r_t(\theta)\textrm{\^{A}}_t,clip(r_t(\theta),1-\epsilon,1+\epsilon)\textrm{\^{A}}_t)]
\label{eq:ppo_clip}
\end{equation}

The comfort zone is defined with the hyperparameter $\epsilon$. $ clip(r_t(\theta),1-\epsilon,1+\epsilon)$ is the clipping function, which clips the probability ratio of the surrogate objective. $\textrm{\^{A}}_t $ is the advantage estimation function at time step $t$. At the end, we take the minimum of the clipped and unclipped objective, so that the final objective is the lower (pessimistic) bound \cite{schulman2017proximal}.

\subsubsection{Concluding Remarks on RL}

In summary, experimental results show that PPO reaches high performance based on a simplistic model, while Q-learning with function approximation is poorly understood and suffers from a lack of robustness \cite{schulman2017proximal}. In this paper we will test this hypothesis and perform a direct comparison of these important DRL methods applied to query optimization.

\section{Architecture}
\label{sec:arch}
In this section we introduce FOOP - a \underline{F}ully \underline{O}bserved \underline{Op}timizer - that uses RL for query optimization. In a first step, we show how to model query optimization as a Markov Decision Process (MDP). Secondly, we discuss how to represent queries and database information as feature vectors that can be used for deep reinforcement learning. The major research challenge is to find a good feature representation as well as the right reinforcement learning algorithm  such that the learning process produces a model which generalizes well and does not get trapped in local optima or never stabilizes. 

\pagebreak

\subsection{Modeling}
To express a problem as an RL problem, we need to formulate it as an MDP, which consists of a five-tuple, as mentioned in Equation \ref{eq:mdp} in Section \ref{sec:bg_rl_mdp}. For the sake of readability, we repeat the specification here:

\begin{equation} 
\langle S,A,P(s,a),R(s,a),S_0 \rangle
\end{equation}

We will now walk through each of these components and describe them with concrete examples based on our sample database shown in Figure \ref{fig:sample_db_er} and the query '$P \bowtie OI \bowtie O \bowtie C$':

\begin{itemize}
\item $S$ (states): The states are all possible (sub)query plans. The complete query plans are the terminal states. Since FOOP is fully observed, we show all involved relations at any time. For our sample query, a sub set of all states is as follows: \\
$S={ [P;OI;O;C],  [(P,OI);O;C], [(OI,P);O;C], ...}$
%\varepsilon , AB,AC,BA,BC,CA,CB,(AB)C,(AC)B,...}$%(BA)C,(BC)A,(CA)B,(CB)A,A(BC),A(CB),B(AC),B(CA),C(AB),C(BA)}$

In the example above, every state is represented by square brackets "[]". For instance, [P;OI;O;C] is the initial state. Each relation that is not joined yet, is separated with a semicolon. The second state is [(P,OI);O;C] where the parentheses "()" indicate a sub query plan.

\item $A$ (actions): The actions are all possible joins included in all query plans. Hence, the total action space is a list with a size larger than $n!$ where $n$ is the number of relations contained in all queries. For our running example, the action space is as follows:

$A = (P \bowtie OI), (OI \bowtie P) , (OI \bowtie O) , (O \bowtie C) ,\\
(C \bowtie O) , (C \bowtie B), ...$

In Section \ref{sec:arch_feat_act} we will discuss a potential simplification to reduce the complexity of the action space.

\item $S_{t+1} \sim P(s,a)$ is the new state you reach when you are in state \textit{s} and take action \textit{a}. Assume \\$ s = [(P,OI);O;C]$ and $a=(C \bowtie O)$. Then the new state is $P(s,a) = [(P,OI);(C,O)]$.

\item $R(s,a)$ is the reward being in state \textit{s} while taking action \textit{a}: The rewards are the negative costs of the resulting query plan. The costs are {\it evaluated only in a terminal state}. In other words, FOOP only receives the costs for the final query plan and not for sub query plans. We introduce the cost model and the reward handling in Section \ref{sec:arch_feat_rew}.

\item $S_0$ for initial state: The initial state is: $[P;OI;O;C]$.

\end{itemize}

\subsection{Featurization}
RL learning algorithms need an environment in which they can interact with the MDP as explained in Section \ref{sec:bg_rl_mdp}. We will now describe all the components of the agent-environment that are necessary to solve a query optimization problem with reinforcement learning. 

\subsubsection{Observation/State}
\label{sec:arch_feat_obs}
An observation represents the state in which the agent currently is. Since we want to provide as much information as possible to our learning algorithms, we formulate query optimization as a fully observed RL problem. Hence, the information of the database and the respective queries have to be encoded in such a way that the representation can be learned by the RL algorithms. The encoded observation serves as input for a neural network (NN) in order to learn from the observations.

\begin{figure*}[ht]
\includegraphics[width=18.5cm]{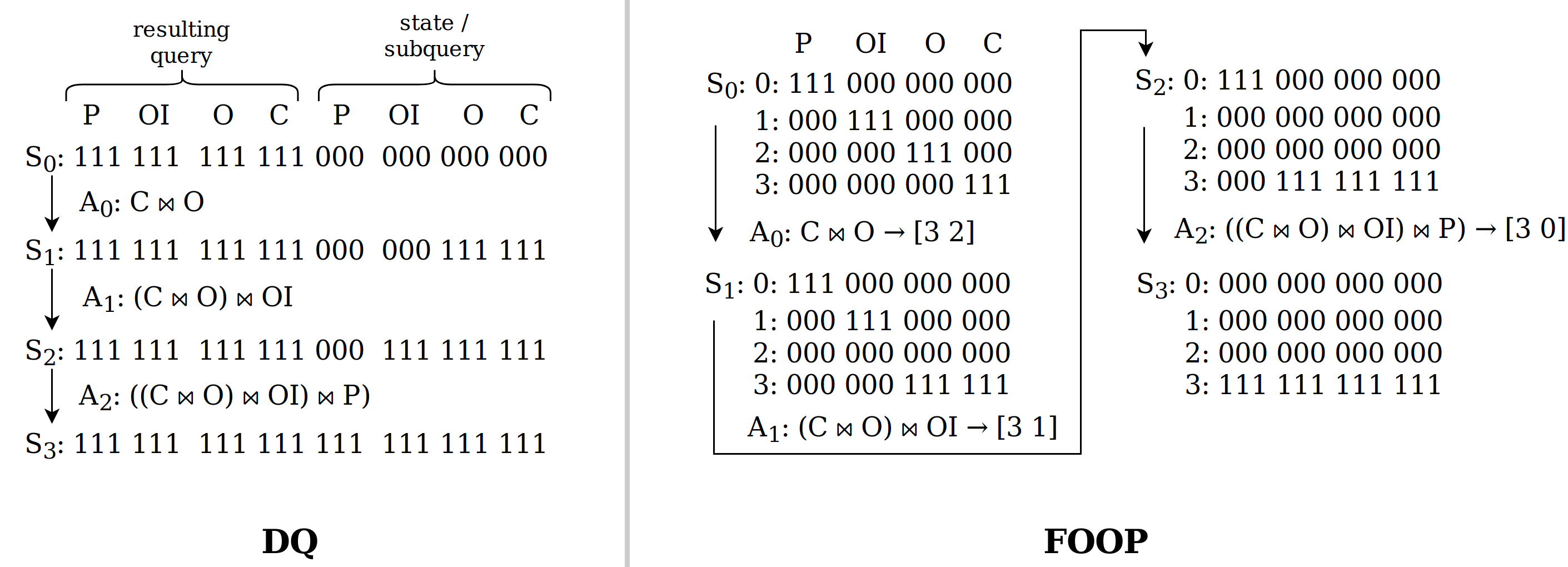}
\caption{Observation featurization in DQ by Krishnan et al. and FOOP presented on the sample query '$P \bowtie OI \bowtie O \bowtie C$' using the sample database shown in Figure \ref{fig:sample_db_er}.}
\label{fig:arch_feat_obs}
\end{figure*}

For the encoding we followed partly the idea of Krishnan et al. \cite{krishnan2018learning}, where each column of the database is used as a single feature. A state is represented as a binary one-hot vector where every digit represents a column of the database. The size of the vector corresponds to the number of all columns over all tables. This vector is used twice (see left example in Figure \ref{fig:arch_feat_obs}): The first vector represents the {\it original query} and marks which columns are involved in the query. Since the query involves all tables, all columns are set to 1. The second vector represents the {\it state of the sub query plan}. In initial state $S_{0}$ all columns are set to zero.

The next step is to perform an action. In this case, action $A_{0}$ suggests to execute the sub query $C \bowtie O$. As a consequence, state $S_{1}$ needs to be updated. In particular, all columns that are involved in this join need to be set to 1 in the second vector. Afterwards action $A_{1}$ is executed and state $S_{2}$ needs to be updated, etc.

Our proposed approach FOOP extends that idea as you can see on the right side in Figure \ref{fig:arch_feat_obs}. Instead of just having a vector that represents the currently joined sub query, we create a symmetric matrix. This matrix represents a table or a sub query in each row. All rows of the matrix together include the needed tables for the final query. With this matrix, we represent the whole database during the process of query execution. 

In the initial state $S_0$, the matrix represents all necessary tables for the query in a separate row. For instance, row 0 has the first three bits set to represent table P. Row 1 has the bits 4 to 6 set, to represent table OI, etc. 

Then action $A_{0}$ performs the sub query $C \bowtie O$, which involves the tables O and C that are represented by the vector [3 2]. 

Next, state $S_{1}$ needs to be updated. We can see that rows 0 and 1 still represent the tables $P$ and $OI$, i.e. the respective columns are set to 1. However, row 3 contains the result of the join $(C \bowtie O)$. Due to the fact that the input vectors of NNs always have to have the same size, we are forced to keep row 2 as an empty row. 

In every further step we add a table to the sub query, until there is just the final query left as presented in the last step $S_3$. 

The advantage of our new featurization approach is as follows: The reinforcement agent knows at any state which sub queries still have to be joined to reach a terminal state. In addition, the agent also sees which sub queries were already joined in the past. 

\subsubsection{Actions}
\label{sec:arch_feat_act}
\begin{figure}[ht]
\includegraphics[width=8.5cm]{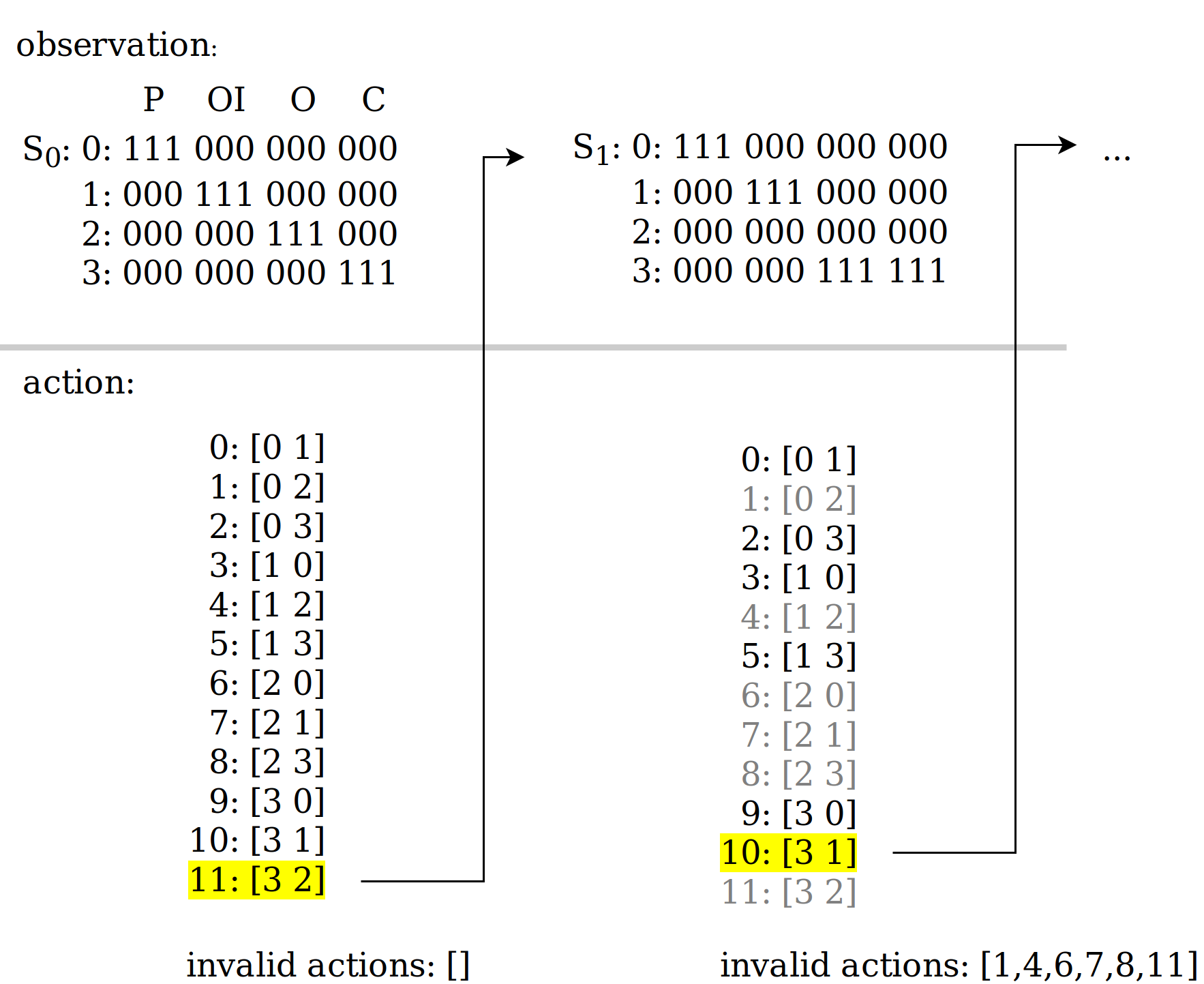}
\caption{Action featurization in FOOP }
\label{fig:arch_feat_act}
\end{figure}

We will now discuss how agents take actions and thus provide details on how we construct the action space.
To create our action space, we can come back to our observation matrix. Every row of the observation matrix is a join candidate, as you can see in state $S_0$ of Figure \ref{fig:arch_feat_act}. To construct the action space, we take all combinations of all join candidates for joining two tables. 

This results  in an action space of $n*(n-1)$, where $n$ is the total number of tables in the database (as shown in the lower half of Figure \ref{fig:arch_feat_act}). For instance, row 0 represents the join $P \bowtie OI$. Row 1 represents the join $P \bowtie O$, etc. In our example, we assume that row 11 (which is highlighted in yellow) is selected by a random agent.

As mentioned in Section \ref{sec:arch_feat_obs} there are empty rows in the observation matrix, during the query optimization process. This means, not all rows in our observation matrix are valid join candidates. Due to that, we have invalid actions (highlighted in light gray) in our action space, as you can see in the right lower part of Figure \ref{fig:arch_feat_act}. All these actions are invalid since they would join row 2, which is an empty row and thus not a join candidate.

Having invalid actions is a common issue in DRL. We solved that issue by creating an {\it action masking layer} in the NN, similar to the approach of Lee et al. for playing autonomously StarCraft II \cite{lee2018modular}. The basic idea is that the output layer of the NN is multiplied with an action mask. This results in a new output, where the values of the invalid actions are set to zero. The action mask has the size of all possible actions (valid actions are represented with a 1 and invalid actions with 0).

\subsubsection{Reward}
\label{sec:arch_feat_rew}
We will now discuss how agents receive a reward when executing an action. A reward is defined as the {\it negative costs of a query plan}. Unfortunately, it would be too time consuming to execute every planed query on a DBMS. Therefore, we need to fall back to cost models of traditional query optimizers. Even though these cost models are sometimes erroneous, they serve as good priors for training a machine learning model. We will now introduce the cost model and show how we integrate it into the agent environment.
%\subsubsubsection{Cost Model}
\begin{itemize}
\item \textbf{Cost Model}: We decided to take the cost model, which was introduced in \cite{leis2015good}:

\begin{equation}
C(Q) = 
\left \{
    \begin{tabular}{l}
    $\tau * |R|$; \\
    ...  $if Q = R$ \\
    $|Q|+C (Q_l)+C (Q_r)$; \\ 
    ...  $if Q = Q_l \bowtie^{hj} Q_r$ \\
    $C(Q_l)+\lambda*|Q_l|*max(\frac{|Q_l \bowtie R|)}{|Q_l|},1)$; \\
    ...  $if Q= Q_l \bowtie^{ij} Q_r,Qr = R$\\
    \end{tabular}
\right \}
\label{eq:costmodel}
\end{equation}

In the Equation \ref{eq:costmodel} above,
\begin{itemize}
\item \textit{R} stands for a base relation.
\item \textit{Q} stands for a (sub)query, where $Q_l$ is the left side of a join and $Q_r$ the right side.
\item \textit{$\tau \leq 1$} is a parameter, which discounts a table scan compared to a join.
\item \textit{$\lambda \geq 1$} is a constant to approximate by how much an index lookup is more expensive than a hash table lookup.
\item \textit{$|\hspace{0.2cm}|$} stands for the cardinality estimation function.
\item \textit{$hj$} and \textit{$ij$} stand for hash join and index nested loop join, respectively.
\end{itemize}
The cost model is tailored for main-memory databases. That means it only measures the number of tuples that pass through each operator and it does not take I/O costs into account. Further, it only distinguishes between hash joins (hj) and index nested loop joins (ij). This cost model is very simplistic compared to cost models of commercial DBMSes. Nevertheless it performs very similar to the cost model of PostgreSQL as Leis et al. \cite{leis2015good} pointed out. We set the constants according to the paper ($\tau=0.2$, $\lambda=2$) \cite{leis2015good}.

%\subsubsubsection{Reward Mapping}
\item \textbf{Reward Mapping}: As mentioned above, the negative costs are the reward for our RL model. The cost model applied on a large database produces cost values in the range of $10^6$ to $10^{18}$.  DRL methods usually operate with reward values in the range (-10,10). Hence, we need to normalize the cost values to a usable range for the RL algorithms. We use a square root function shown in Equation \ref{eq:rewardmapping} to normalize the cost values. Linear normalization did not work well, due to the distribution of the cost values.

\begin{equation}
N(Q) = \frac{\sqrt{C(Q)}}{\sqrt{upperbound}}*-10
\label{eq:rewardmapping}
\end{equation}

Since good query plans are just a small fraction of the whole space of all possible query plans, we {\it clip our reward space}. In Figure \ref{fig:apply_rewardmapping} you see the mapping of the cost values to the reward. The square root function is applied on all cost values lower than $10^{13}$, with $10^{13}$ as the upperbound. The cost values bigger than $10^{13}$ are clipped to the min reward of $-10$.

\begin{figure}[ht]
\includegraphics[width=8.5cm]{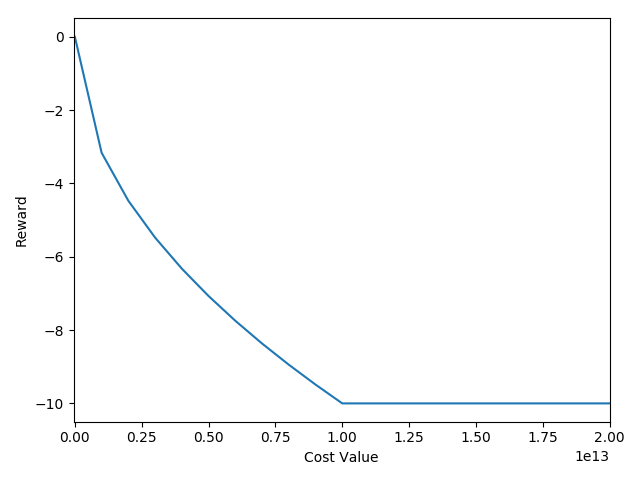}
\caption{Mapping of the costs values to the reward scale (-10,0).}
\label{fig:apply_rewardmapping}
\end{figure}
\end{itemize}
\section{Evaluation}
\label{sec:eval}
In this section, we present and discuss the results of the experiments using various reinforcement learning techniques for query optimization. In particular, we will address the following research questions:

\begin{itemize}
\item How effective are Q-learning and policy gradient-based reinforcement learning approaches for query optimization?
\item Which of these approaches shows the best performance?
\item Can we even further improve the above mentioned approaches by applying ensemble learning techniques?
\end{itemize}

\subsection{Experimental Setup}
All experiments were performed on a laptop running Ubun- tu version 18.04.1 with a 4-core Intel Core i7 8650U CPU (1.90-4.20 GHz), 16 GB of RAM and no GPU module.

We implemented our approach\footnote{The source code of our approach can be found at https://github.com/heitzjon/mt-join-query-optimization-with-drl} as an extension of gym from OpenAI \cite{openai2019openai} in Python. This allows using the RL interface from OpenAI and the baseline RL algorithms provided by Ray RLLib \cite{rayrllib2019rayrllib}. The NN models are written with Tensorflow \cite{tensorflow2019tensorflow}. Our query optimizer is located in the application layer, so we are independent from the DBMS. To calculate the expected costs as explained in Section \ref{sec:arch_feat_rew} we use the cardinality estimator from PostgreSQL 10.6 \cite{postgresql2019postgresql}. Furthermore, for our end-to-end performance experiment we use PostgreSQL 10.6 \cite{postgresql2019postgresql} as the query execution engine. 

For the evaluation we use the Join Order Benchmark (JOB) introduced by Leis et al. \cite{leis2015good}. The benchmark consists of queries with 3 to 16 joins, with an average of 8 joins per query. There are 33 different query structures. Each structure exists in 2 to 6 different variants, which results in a query set of 113 queries. All queries are realistic in a sense that they answer a question of a movie enthusiast and are not constructed to 'trick' a query optimizer \cite{leis2015good}.

JOB is based on the IMDB data set \cite{imdb2019imdb}, which is freely available. The IMDB is a real-world data set consisting of 21 tables. It contains information about movies and movie related facts like actors, directors etc. We use the same snapshot from May 2013 as Leis et al. \cite{leis2015good} do. The data set comprises 3.6 GB of highly correlated and non-uniformly distributed data. The largest tables are: $cast\_info$ with 36 million rows and $movie\_info$ with 15 million rows \cite{leis2015good}.

If not further specified, all results are presented on all 113 JOB queries. To ensure that the performance is evaluated only on queries which have not been seen by the model during training, we use the 4-fold cross validation introduced by Krishnan et al. \cite{krishnan2018learning}. Each of the 113 queries occurs at least in one test set. The train and test sets consist of 80 and 33 queries, respectively. 

\subsection{Evaluation of RL algorithms}
\label{sec:eval_rl}
In the first part of the evaluation we will analyze the performance of different RL algorithms. We will start with {\it deep Q-networks} (DQN) and discuss some enhancements. Afterwards we will analyze {\it proximal policy optimization} (PPO).

\subsubsection{Deep-Q-Networks (DQN)}

Let us start with the evaluation of the DQNs, which we introduced in Section \ref{sec:bg_rl_ql}. We begin with the vanilla DQN, which was used in the DQ paper \cite{krishnan2018learning} and that we have implemented and configured according to the information provided in the paper.

The most important hyper-parameter values of the vanilla DQN can be found in Table \ref{tbl:configVanillaDQN}. The full set of parameter values can be found on our github repository\footnote{https://github.com/heitzjon/mt-join-query-optimization-with-drl/blob/master/agents/run/configs.py}. The vanilla DQN is trained over 4000 iterations. The learning starts after time step 1000 and the target network gets updated every 500 time steps. For this model we use 2-step Q-learning and a neural network with two hidden layers with 256 neurons each.  

As introduced in Section \ref{sec:arch_feat_rew}, our DQN uses the expected costs calculated by the cost model Equation \ref{eq:costmodel} as the negative reward for each query. We only give the DQN a reward signal after planing the full query, to reach a stable learning process and to give the DQN a chance to learn. 

\begin{center}
\begin{tabular}{ |p{5.5cm}|p{2cm}|  }
\hline
%\multicolumn{2}{|c|}{\textbf{Vanilla DQN Configuration}} \\
Parameter & Value \\
\hline
Training iterations: & 4000\\
Learning starts (in time steps): & 1000\\
Target network update (in time steps): & 500\\
n-step Q-learning: & 2 \\
Hidden layers: & [256 256]\\
\hline
\end{tabular}
\captionof{table}{Hyper-parameter values of Vanilla DQN network}
\label{tbl:configVanillaDQN} 
\end{center}

\begin{figure}[ht]
\includegraphics[width=8.5cm]{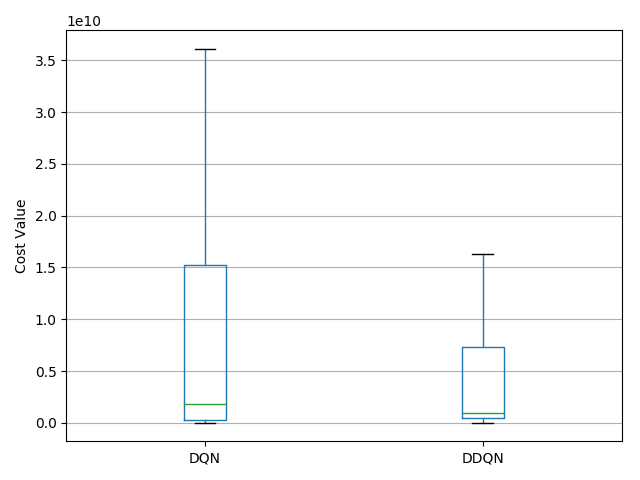}
\caption{Cost values for optimizing 113 queries using reinforcement learning algorithms. The figure compares a vanilla deep-Q-network (DQN) with a double deep-Q-network (DDQN) using priority replay.}
\label{fig:eval_dqn}
\end{figure}

The results on the left side of Figure \ref{fig:eval_dqn} show the estimated cost values trained on 80 queries and tested on 33 queries. The cost value is computed based on the cost model shown in Section 4, see Equation \ref{eq:costmodel}.

%{\bf TODO Kurt: Check again: Do you compare the predicted values against the costs of the Postgres cost model? $->$ No I don't. KS: How do we evaluate the accuracy of these costs if we don't compare them to some kind of baseline? $->$ Do you want to test if our cost model delivers accurate costs? Then the answer is we can't but the paper \cite{leis2015good}, did a comparison with it. Or do you want to compare the costs of the optimized query plans? That is discussed in section 5.4.} 

%{\bf TODO Jonas: We need to verify that the output of DQN is good. Assuming that DQN produced an optimal query plan for a given query, how do we verify that this is actually the best plan? $->$  We wouldn't have to use RL if we would know which query plan the perfect  one is (np-hard problem). We compare our results in section 5.4 against an other query optimization technique. Further infernally we grade our generated query plans with the cost model. This allows us to compare the different query plans with each other.}

We can see that the average cost value for the deep-Q-network (DQN) is around 0.2*1e10  (see left side of Figure~\ref{fig:eval_dqn}). However, the minimal and maximal value range between 0.1*1e10 and 3.5*1e10, which indicates a high variance in the estimated query costs resulting in expensive query plans for about half of the observed queries.

In order to reduce the high variance of the estimated costs and to reach a more stable optimization solution, we extended the vanilla DQN. In particular, we extended the vanilla DQN model with {\it double DQN} and {\it priority replay}. In addition, we used a larger neural network (NN) to achieve a higher abstraction level for finding an optimal Q-function. The configuration of the DDQN is listed in Table \ref{tbl:doubleDeepQ}. The DDQN is trained over 40,000 iterations. The learning starts after time step 160,000 and the target network gets updated every 32,000 time steps. For this model we use 2-step Q-learning as well and a neural network with two hidden layers of 6,272 and 1,568 neurons each.  

The results of the {\it double deep-Q-network} (DDQN) with priority replay are shown on the right side in Figure \ref{fig:eval_dqn}.  

\begin{center}
\begin{tabular}{ |p{5.5cm}|p{2cm}|  }
\hline
Parameter & Value \\
\hline
Training iterations: & 40,000\\
Learning starts (in time steps): & 160,000\\
Target network update (in time steps): & 32,000\\
n-step Q-learning: & 2 \\
Hidden layers: & [6272 1568]\\
\hline
\end{tabular}
\captionof{table}{Hyper-parameter values of the double deep-Q-network}
\label{tbl:doubleDeepQ} 
\end{center}

Figure \ref{fig:eval_dqn} shows the cost spread off all 113 queries. Even though the median cost value (green line) of DQN and DDQN is very similar, the inter-quartile range (blue box) of the vanilla DQN is by a factor of two larger than the one of the DDQN. In addition, the maximum cost value of the vanilla DQN  is by a factor of two larger than the maximum of the DDQN. The minimum values of the vanilla DQN and DDQN are similar. In short, the Q-function of DDQN produces far less expensive query plans than the vanilla DQN.

\pagebreak

\subsubsection{Proximal Policy Optimization (PPO)}
In the second step we evaluate proximal policty optimization (PPO), which is used in ReJoin \cite{marcus2018deep}. We have implemented that approach using the reinforcement learning algorithms provided by RLLib of Ray\footnote{https://github.com/ray-project/ray}. The basics of PPO are introduced in Section \ref{sec:bg_rl_ppo}. Unfortunately, the ReJoin \cite{marcus2018deep} paper does not provide the configuration of the used PPO model. Due to that we created and tuned our own configuration, which you can find in Table \ref{tbl:PPO}. The full set of parameter values can be found on our github repository\footnote{https://github.com/heitzjon/mt-join-query-optimization-with-drl/blob/master/agents/run/configs.py}. The PPO is trained over 200,000 iterations. During the training process we clip policies with a bigger deviation than 0.3. Furthermore, we us for our model a neural network with two hidden layers of 256 neurons. 

\begin{center}
\begin{tabular}{ |p{5.5cm}|p{2cm}|  }
\hline
Parameter & Value \\
\hline
Training iterations: & 200,000\\
Clipping coefficient ($\epsilon$): & 0.3\\
Hidden layers: & [256 256]\\
\hline
\end{tabular}
\captionof{table}{Hyper-parameter values of the PPO network}
\label{tbl:PPO} 
\end{center}

\begin{figure*}[ht]
\includegraphics[width=19cm]{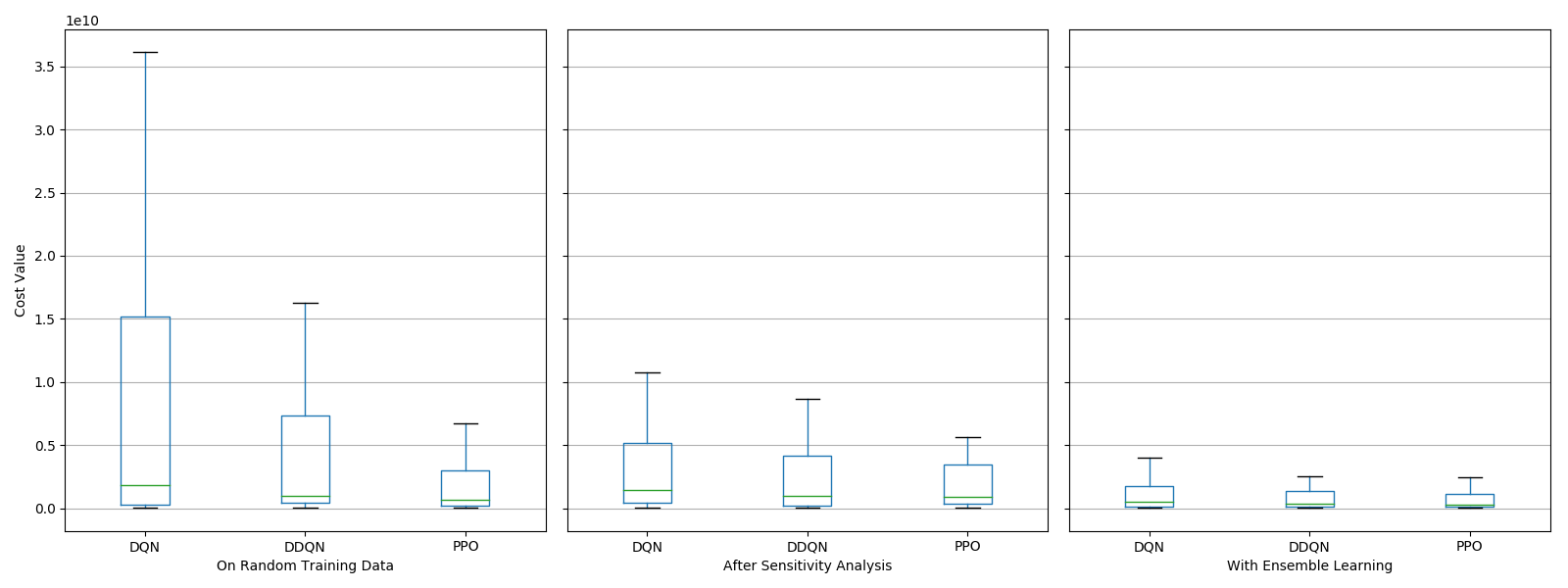}
\caption{Comparing three different DRL algorithms in FOOP a) on a random training data, b) after adapting the training and test data sets and c) with Ensemble Learning}
\label{fig:eval_all}
\end{figure*}

PPO outperforms both DQN configurations, as presented on the left side of Figure \ref{fig:eval_all}. PPO is able to reduce the inter-quartile range and the maximum by more than a factor of two compared to DDQN.
%\begin{figure}[ht]
%\includegraphics[width=8.5cm]{imgs/compare_dqn_ddqn_ppo.png}
%\caption{Comparing DRL algorithms in FOOP on %a random training data }
%\label{fig:eval_all}
%\end{figure}

\subsection{Enhance RL models}
In this section we will to evaluate how we can improve the DRL models. First, we will discuss how the split of training and test data affects the learning process. Afterwards we introduce {\it ensemble learning} to improve the DRL-based query optimizer.

\subsubsection{Re-Arranging Training and Test Data Sets}
In our previously presented experiments, we used a random data split for generating training and test sets. As Krishnan et al. \cite{krishnan2018learning} pointed out, it is important that the training queries cover all relations and use all important relationships in the database. So we introduce a new training/test data split, which can be found in our github repository\footnote{https://github.com/heitzjon/mt-join-query-optimization-with-drl/tree/master/agents/queries/crossval\_sens}. The key requirements for the data split are:

\begin{itemize}
\item All relations have to be represented in every training set.
\item The training sets have to contain all possible join conditions.
\item To have accurate test conditions, every test set has to contain the largest possible variety of different join queries.
\end{itemize}

We ran our experiments with the new data split for the training and test set. The results are shown in the middle of Figure \ref{fig:eval_all}. As we can see, the new arrangement of the training and test data improves the query optimization performance of all models. Especially the inter-quartile range and the maximum of DQN and DDQN can be reduced significantly. Nevertheless, PPO still outperforms the DQN models. The better performance can be explained by the new training sets, which cover all relations and join conditions of the database. This enables the DRL-based optimizer to reduce overfitting of the Q-function and policy-function.

%\begin{figure}[ht]
%\includegraphics[width=8.5cm]{imgs/compare_dqn_ddqn_ppo_sens.png}
%\caption{Comparing three different DRL algorithms after adapting the training and test data sets.}
%\label{fig:eval_all_sens}
%\end{figure}

\subsubsection{Ensemble Learning}
In RL the training process is a stochastic process. This means that the learned policies from different training runs can have divergences resulting in different optimal query plans for the same query. {\it Ensemble learning} is a method, which helps us to transform that issue into a benefit. Ensemble learning is often used to combine multiple learning algorithms, with the aim to create better predictions than a single learning algorithm. In other words, the agent asks for a second or a third opinion before taking a decision \cite{polikar2006ensemble}.

%\clearpage
To apply ensemble learning, we first train every DRL model five times and thus receive five different learned policies. Afterwards use all five learned policies to optimize a given query. As a result, we will receive five different "optimal" query plans. Finally, we chose the query plan with the lowest costs. 

%$->$ We do not execute the same policy five times. We train five different policies from the same DRL model. It results in five slightly different policies, since the learning process is a stochastic process. -> Good ;-)

%Each policy creates a query plan and the cost model returns a cost estimate for each query plan. As the final step we take the query plan with the lowest cost value as the query plan to execute.

The results of using three different DRL-based query optimization with ensemble learning are shown on the right side of Figure \ref{fig:eval_all}.  We are able to reduce the inter-quartile range and the maximum of all models significantly. In addition, we were even able to lower the median of all three models. It is not surprising, that we get better results, since we take just the best query plans from five different learned policies. Moreover, we could also reduce the amount of outliers. 

%\begin{figure}
%\includegraphics[width=8.5cm]{imgs/compare_dqn_ddqn_ppo_ensemble.png}
%\caption{Comparing DRL algorithms in FOOP %with Ensemble Learning}
%\label{fig:eval_all_el}
%\end{figure}

\subsection{Comparison to other Approaches}
So far we compared the DRL algorithms introduced in ReJoin \cite{marcus2018deep} and DQ \cite{krishnan2018learning}. However, for DQ we used the featurization of FOOP. The advantage of FOOP's featurization is that the learning algorithms have more information about the query execution process resulting in a smaller action space.

Now we will compare these approaches to a traditional query optimization approach based on dynamic programming to better understand the advantages and disadvantage of DRL-based approaches.

%Later we will discuss whether we can compare the results of FOOP to the results of ReJoin \cite{marcus2018deep} and DQ \cite{krishnan2018learning} directly.

\subsubsection{Dynamic Programming with Left-Deep Plans}
DP algorithms were introduced to query optimization in System R \cite{astrahan1976system} and are still widely used in commercial DBMSes. The basic assumption of these algorithms is that the {\it optimal sub query} plan is part of the {\it optimal query} plan. The algorithm therefore compares sub queries, which have the same outcome, and picks the sub query with the lowest costs to proceed. However, this process is very memory- and computing-intensive, especially for large queries. 

Typical approaches use a recursive process which grows on a lower bound of $n!$ where $n$ represents the number of relations in the query. To reduce that complexity, System R \cite{astrahan1976system} introduces rules like 'left-deep query trees only' and 'no cross joins'. We implemented the bottom-up DP algorithm from System R \cite{astrahan1976system} with restrictions to left-deep trees. The algorithm can be found in our github repository. Note that this approach is still very compute-intensive. Hence, DBMSes like PostgreSQL limit the size of queries processed by dynamic program (e.g. 8-way-joins). When queries contain more than 8 relations to join, the optimizer greedily selects  the remaining relations or uses genetic algorithms for optimization. 

\begin{figure}
\includegraphics[width=8.5cm]{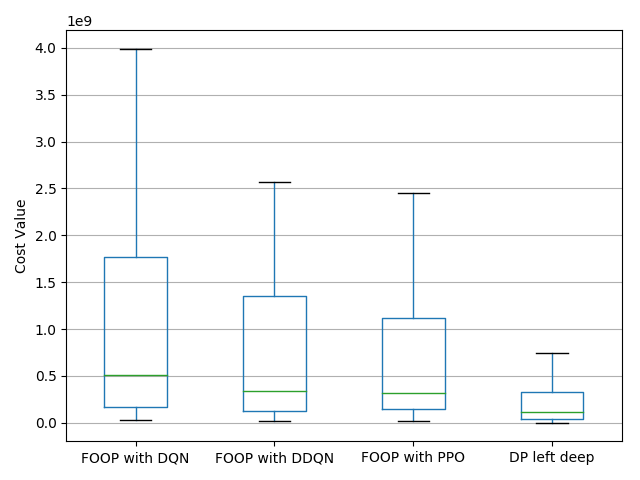}
\caption{Comparison of FOOP to DP left-deep }
\label{fig:eval_dp_foop}
\end{figure}

In Figure \ref{fig:eval_dp_foop} we compare the three previously discussed DRL-based optimization strategies against dynamic programming (DP). As we can see, DP outperforms the DRL-based approaches. However, the optimization run time complexity for, e.g. PPO, is much better than for dynamic programming as we will show in Section \ref{sec:training_time}. 

\begin{figure}
\includegraphics[width=8.5cm]{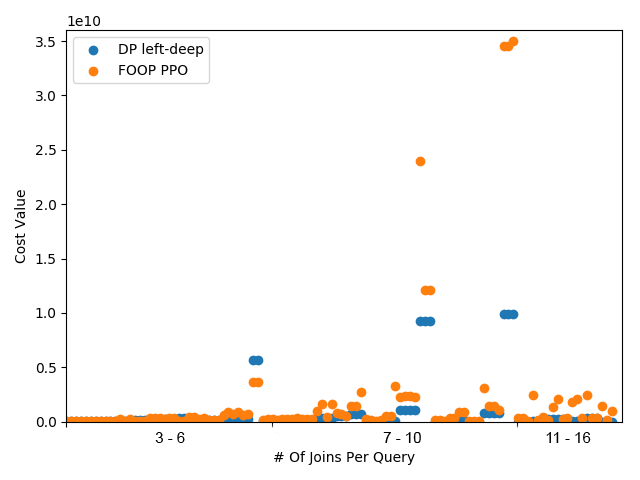}
\caption{Query-by-query comparison of FOOP with PPO and DP left-deep}
\label{fig:eval_dp_foop2}
\end{figure}

Let us now take a closer look at PPO when compared to dynamic programming. In particular, we will analyze the join complexity of the queries. The x-axes of Figure \ref{fig:eval_dp_foop2} shows the queries sorted by number of joins ranging from 3 to 16. We can see that for a large number of queries PPO even outperforms dynamic programming. However, PPO has about 6 outliers that decrease the overall performance. We will now analyze these outliers in more detail.

\subsubsection{Outlier Analysis}
\begin{figure}
\includegraphics[width=8.5cm]{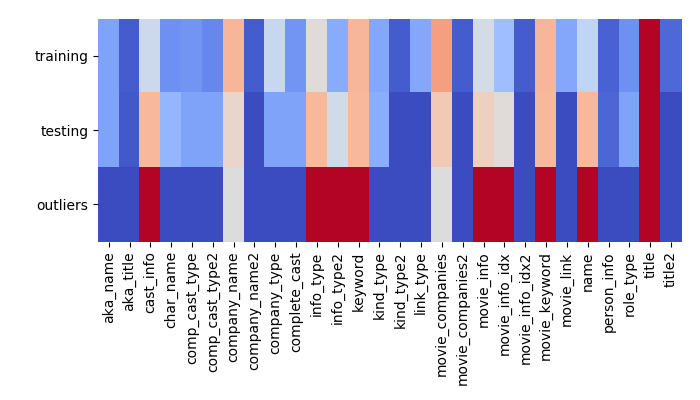}
\caption{The heatmap to shows how often particular tables are accessed by the queries in the training set, test set and the outliers.}
\label{fig:eval_outliers}
\end{figure}

During the analysis of the outliers we observed that all 6 outlier queries come from two different query templates of the JOB benchmark. In particular, the outlier queries belong to the templates 25 and 31. These outliers can be explained by the occurrence of the different tables in the training set versus the test data set. Figure \ref{fig:eval_outliers} shows the occurrence of the tables as a heatmap. The color red indicates that the respective table is heavily used, i.e. it is "hot", while the color blue indicates that the respective table is hardly used, i.e. it is "cold".  We can observe that for outlier queries there is a higher number of "hot" tables in the test data, while the respective tables in the training data tend to be "colder". In other words, the test data does not contain enough training examples and hence these queries could not be learned well.

\subsection{Training Time}
\label{sec:training_time}
In contrast to traditional query optimizers, DRL-based query optimizers initially need to be trained before they can be used. In this section we describe the training latency of the different DRL-models used in FOOP.
The training process of vanilla DQN over 4,000 time steps and  DDQN over 40,000 time steps takes about 7 minutes each. The PPO model is trained over 200,000 time steps and takes roughly 25 minutes.

The training process in FOOP with ensemble learning increases by a factor of five, since we train each model five times. The DQN models take 45 minutes and the PPO model even needs 125 minutes to complete the training process. In comparison, ReJoin needs 3 hours to complete the training process on a slightly smaller hardware configuration. 

\begin{figure}[ht]
\includegraphics[width=8.5cm]{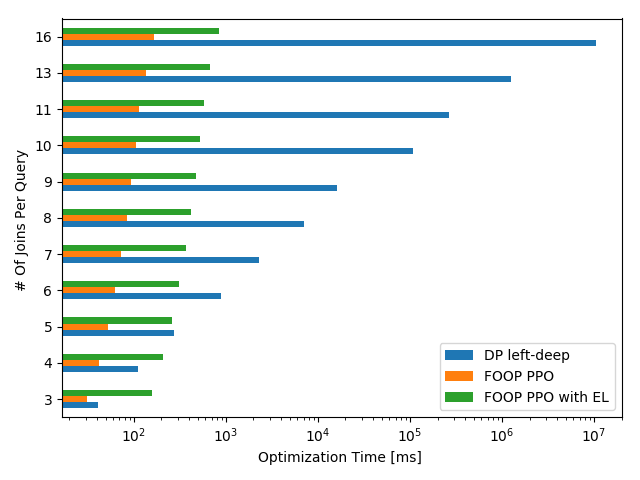}
\caption{Comparing optimization latency of FOOP with DP left-deep DP. }
\label{fig:eval_planning_latency}
\end{figure}

One of the biggest advantage of query optimization with DRL is, that the optimization latency is linear to the used relations of a query \cite{marcus2018deep}, which results in a complexity of $O(n)$. The latency grows linear with the join size, even if we use ensemble learning to improve our query plans, as presented in Figure \ref{fig:eval_planning_latency}. Left-deep DP, on the other hand, has an optimization latency which is factorial to the relations of the query, this corresponds to the complexity of $O(n!)$. 

%{\bf TODO: Since all queries in JOB are optimized exactly, the results could be shown as  $C_{FOOP}/C_{EXACT}$ $->$ I wasn't aware of that where did you find the exactly optimized JOB queries? In which graph would you like to show that? TODO2: This was a comment by Michael Grossniklaus. I will ask him again}

\section{ Conclusion and Limitation }
\label{sec:conc}
In this paper we analyzed various deep reinforcement learning algorithms for query optimization and introduced FOOP - a deep reinforcement learning-based Fully Observed Optimizer. FOOP is currently limited to join order optimization but can easily be extended to optimize select-project-join queries or even non-relational operators as presented in DQ \cite{krishnan2018learning}. FOOP is implemented in an RL-friendly environment which enables experimenting with the most cutting-edge RL algorithms.

Our experimental evaluation shows that Proximal Policy Optimization (PPO) reinforcement learning algorithms outperform Deep Q-Networks (DQN) in the task of join order optimization. We also suggest to use ensemble learning in combination with DRL to mitigate stability issues of the stochastic training process. The results demonstrate that ensemble learning significantly improves the overall performance of the RL algorithms. FOOP produces query plans with cost estimations in a similar range to traditional query optimizers based on left-deep dynamic programming algorithms. However, the optimization latency of FOOP is significantly lower than for left-deep DP algorithms.

Based on our experiments, we see the following avenues of further research:

\begin{itemize}
\item One of the disadvantages of deep reinforcement learning is that pre-training certain models is very compute-intensive. One solution to overcome this problem is to integrate bootstrapped expert knowledge into the pre-training phase for faster learning.
\item  Some of the cost models of traditional query optimizers suffer from erroneous cardinality estimations. To solve this problem, one could add a fine-tuning step to the training process to reuse cardinality data from already executed queries. 
\item In order to provide an end-to-end query optimizer, FOOP needs to be extended with further functions like selections, projections as well as aggregations.  
\end{itemize}

\section*{Acknowledgement}
We would like to thank Joseph Hellerstein and Zongheng Yang for the fruit-full discussions about DRL in query optimization. The advice given by Michael Grossniklaus about query optimization and Paul Bertucci about database tuning has been a great help. We also want to thank Katharina Rombach for the many discussions about reinforcement learning.

\bibliographystyle{plain}
\bibliography{Manuscript}

\begin{thebibliography}{10}

\bibitem{akdere2012learning}
Mert Akdere, Ugur {\c{C}}etintemel, Matteo Riondato, Eli Upfal, and Stanley~B
  Zdonik.
\newblock Learning-based query performance modeling and prediction.
\newblock In {\em 2012 IEEE 28th International Conference on Data Engineering},
  pages 390--401. IEEE, 2012.

\bibitem{astrahan1976system}
Morton~M. Astrahan, Mike~W. Blasgen, Donald~D. Chamberlin, Kapali~P. Eswaran,
  Jim~N Gray, Patricia~P. Griffiths, W~Frank King, Raymond~A. Lorie, Paul~R.
  McJones, James~W. Mehl, et~al.
\newblock System r: relational approach to database management.
\newblock {\em ACM Transactions on Database Systems (TODS)}, 1(2):97--137,
  1976.

\bibitem{avnur2000eddies}
Ron Avnur and Joseph~M Hellerstein.
\newblock Eddies: Continuously adaptive query processing.
\newblock In {\em ACM sigmod record}, volume~29, pages 261--272. ACM, 2000.

\bibitem{crepinvsek2009efficient}
Matej Crepin{\v{s}}ek and Luka Mernik.
\newblock An efficient representation for solving catalan number related
  problems.
\newblock {\em International journal of pure and applied mathematics},
  56(4):598--604, 2009.

\bibitem{ganapathi2009predicting}
Archana Ganapathi, Harumi Kuno, Umeshwar Dayal, Janet~L Wiener, Armando Fox,
  Michael Jordan, and David Patterson.
\newblock Predicting multiple metrics for queries: Better decisions enabled by
  machine learning.
\newblock In {\em 2009 IEEE 25th International Conference on Data Engineering},
  pages 592--603. IEEE, 2009.

\bibitem{goodfellow2016deep}
Ian Goodfellow, Yoshua Bengio, Aaron Courville, and Yoshua Bengio.
\newblock {\em Deep learning}, volume~1.
\newblock MIT press Cambridge, 2016.

\bibitem{tensorflow2019tensorflow}
{Google Brain Team}.
\newblock Tensorflow: An end-to-end open source machine learning platform.
\newblock {\em https://www.tensorflow.org/}, June 2019.

\bibitem{graefe1991volcano}
Goetz Graefe and William McKenna.
\newblock The volcano optimizer generator.
\newblock Technical report, COLORADO UNIV AT BOULDER DEPT OF COMPUTER SCIENCE,
  1991.

\bibitem{vt22019vt2}
Jonas Heitz and Kurt Stockinger.
\newblock Learning cost model: Query optimization meets deep learning.
\newblock {\em Zurich University of Applied Science, Project Thesis}, 2019.

\bibitem{hellerstein2005readings}
Joseph~M Hellerstein and Michael Stonebraker.
\newblock {\em Readings in database systems}.
\newblock MIT Press, 2005.

\bibitem{hessel2018rainbow}
Matteo Hessel, Joseph Modayil, Hado Van~Hasselt, Tom Schaul, Georg Ostrovski,
  Will Dabney, Dan Horgan, Bilal Piot, Mohammad Azar, and David Silver.
\newblock Rainbow: Combining improvements in deep reinforcement learning.
\newblock In {\em Thirty-Second AAAI Conference on Artificial Intelligence},
  2018.

\bibitem{imdb2019imdb}
IMDB Inc.
\newblock Imdb.
\newblock {\em https://www.imdb.com/interfaces/}, 2019.

\bibitem{ioannidis1996query}
Yannis~E Ioannidis.
\newblock Query optimization.
\newblock {\em ACM Computing Surveys (CSUR)}, 28(1):121--123, 1996.

\bibitem{krishnan2018learning}
Sanjay Krishnan, Zongheng Yang, Ken Goldberg, Joseph Hellerstein, and Ion
  Stoica.
\newblock Learning to optimize join queries with deep reinforcement learning.
\newblock {\em arXiv preprint arXiv:1808.03196}, 2018.

\bibitem{lee2018modular}
Dennis Lee, Haoran Tang, Jeffrey~O Zhang, Huazhe Xu, Trevor Darrell, and Pieter
  Abbeel.
\newblock Modular architecture for starcraft ii with deep reinforcement
  learning.
\newblock In {\em Fourteenth Artificial Intelligence and Interactive Digital
  Entertainment Conference}, 2018.

\bibitem{leis2015good}
Viktor Leis, Andrey Gubichev, Atanas Mirchev, Peter Boncz, Alfons Kemper, and
  Thomas Neumann.
\newblock How good are query optimizers, really?
\newblock {\em Proceedings of the VLDB Endowment}, 9(3):204--215, 2015.

\bibitem{li2012robust}
Jiexing Li, Arnd~Christian K{\"o}nig, Vivek Narasayya, and Surajit Chaudhuri.
\newblock Robust estimation of resource consumption for sql queries using
  statistical techniques.
\newblock {\em Proceedings of the VLDB Endowment}, 5(11):1555--1566, 2012.

\bibitem{lohman2014query}
Guy Lohman.
\newblock Is query optimization a “solved” problem.
\newblock In {\em Proc. Workshop on Database Query Optimization}, page~13.
  Oregon Graduate Center Comp. Sci. Tech. Rep, 2014.

\bibitem{marcus2019neo}
Ryan Marcus, Parimarjan Negi, Hongzi Mao, Chi Zhang, Mohammad Alizadeh, Tim
  Kraska, Olga Papaemmanouil, and Nesime Tatbul.
\newblock Neo: A learned query optimizer.
\newblock {\em arXiv preprint arXiv:1904.03711}, 2019.

\bibitem{marcus2016wisedb}
Ryan Marcus and Olga Papaemmanouil.
\newblock Wisedb: a learning-based workload management advisor for cloud
  databases.
\newblock {\em Proceedings of the VLDB Endowment}, 9(10):780--791, 2016.

\bibitem{marcus2018deep}
Ryan Marcus and Olga Papaemmanouil.
\newblock Deep reinforcement learning for join order enumeration.
\newblock In {\em Proceedings of the First International Workshop on Exploiting
  Artificial Intelligence Techniques for Data Management}, page~3. ACM, 2018.

\bibitem{marcus2018towards}
Ryan Marcus and Olga Papaemmanouil.
\newblock Towards a hands-free query optimizer through deep learning.
\newblock {\em arXiv preprint arXiv:1809.10212}, 2018.

\bibitem{markl2004robust}
Volker Markl, Vijayshankar Raman, David Simmen, Guy Lohman, Hamid Pirahesh, and
  Miso Cilimdzic.
\newblock Robust query processing through progressive optimization.
\newblock In {\em Proceedings of the 2004 ACM SIGMOD international conference
  on Management of data}, pages 659--670. ACM, 2004.

\bibitem{mnih2013playing}
Volodymyr Mnih, Koray Kavukcuoglu, David Silver, Alex Graves, Ioannis
  Antonoglou, Daan Wierstra, and Martin Riedmiller.
\newblock Playing atari with deep reinforcement learning.
\newblock {\em arXiv preprint arXiv:1312.5602}, 2013.

\bibitem{mnih2015human}
Volodymyr Mnih, Koray Kavukcuoglu, David Silver, Andrei~A Rusu, Joel Veness,
  Marc~G Bellemare, Alex Graves, Martin Riedmiller, Andreas~K Fidjeland, Georg
  Ostrovski, et~al.
\newblock Human-level control through deep reinforcement learning.
\newblock {\em Nature}, 518(7540):529, 2015.

\bibitem{openai2019openai}
{OpenAI}.
\newblock Gym.
\newblock {\em https://gym.openai.com}, June 2019.

\bibitem{ortiz2018learning}
Jennifer Ortiz, Magdalena Balazinska, Johannes Gehrke, and S~Sathiya Keerthi.
\newblock Learning state representations for query optimization with deep
  reinforcement learning.
\newblock {\em arXiv preprint arXiv:1803.08604}, 2018.

\bibitem{polikar2006ensemble}
Robi Polikar.
\newblock Ensemble based systems in decision making.
\newblock {\em IEEE Circuits and systems magazine}, 6(3):21--45, 2006.

\bibitem{schaul2015prioritized}
Tom Schaul, John Quan, Ioannis Antonoglou, and David Silver.
\newblock Prioritized experience replay.
\newblock {\em arXiv preprint arXiv:1511.05952}, 2015.

\bibitem{schulman2017proximal}
John Schulman, Filip Wolski, Prafulla Dhariwal, Alec Radford, and Oleg Klimov.
\newblock Proximal policy optimization algorithms.
\newblock {\em arXiv preprint arXiv:1707.06347}, 2017.

\bibitem{silver2016mastering}
David Silver, Aja Huang, Chris~J Maddison, Arthur Guez, Laurent Sifre, George
  Van Den~Driessche, Julian Schrittwieser, Ioannis Antonoglou, Veda
  Panneershelvam, Marc Lanctot, et~al.
\newblock Mastering the game of go with deep neural networks and tree search.
\newblock {\em nature}, 529(7587):484, 2016.

\bibitem{stillger2001leo}
Michael Stillger, Guy~M Lohman, Volker Markl, and Mokhtar Kandil.
\newblock Leo-db2's learning optimizer.
\newblock In {\em VLDB}, volume~1, pages 19--28, 2001.

\bibitem{sutton2018reinforcement}
Richard~S Sutton and Andrew~G Barto.
\newblock {\em Reinforcement learning: An introduction}.
\newblock MIT press, 2018.

\bibitem{postgresql2019postgresql}
{The PostgreSQL Global Development Group}.
\newblock Postgresql: The world's most advanced open source relational
  database.
\newblock {\em https://www.postgresql.org/}, January 2019.

\bibitem{rayrllib2019rayrllib}
{The Ray Team}.
\newblock Rllib: Scalable reinforcement learning.
\newblock {\em https://ray.readthedocs.io/en/latest/rllib.html}, June 2019.

\bibitem{trummer2018skinnerdb}
Immanuel Trummer, Samuel Moseley, Deepak Maram, Saehan Jo, and Joseph
  Antonakakis.
\newblock Skinnerdb: regret-bounded query evaluation via reinforcement
  learning.
\newblock {\em Proceedings of the VLDB Endowment}, 11(12):2074--2077, 2018.

\bibitem{wang2016database}
Wei Wang, Meihui Zhang, Gang Chen, HV~Jagadish, Beng~Chin Ooi, and Kian-Lee
  Tan.
\newblock Database meets deep learning: challenges and opportunities.
\newblock {\em ACM SIGMOD Record}, 45(2):17--22, 2016.

\bibitem{wu2013towards}
Wentao Wu, Yun Chi, Hakan Hac{\'\i}g{\"u}m{\"u}{\c{s}}, and Jeffrey~F Naughton.
\newblock Towards predicting query execution time for concurrent and dynamic
  database workloads.
\newblock {\em Proceedings of the VLDB Endowment}, 6(10):925--936, 2013.

\bibitem{yusufoglu2014neural}
Elif~Ezgi Yusufoglu, Murat Ayyildiz, and Ensar Gul.
\newblock Neural network-based approaches for predicting query response times.
\newblock In {\em 2014 International Conference on Data Science and Advanced
  Analytics (DSAA)}, pages 491--497. IEEE, 2014.

\end{thebibliography}
\end{document}